# Direct Observation of Spectroscopic Inhomogeneities on $La_{0.7}Sr_{0.3}MnO_3$ Thin Films by Scanning Tunnelling Spectroscopy

R. Di Capua[1], C.A. Perroni[1,3], V. Cataudella[1], F. Miletto Granozio[1], P. Perna[2], M. Salluzzo[1], U. Scotti Di Uccio[2], and R. Vaglio[1]

[1] *COHERENTIA - INFM and Dipartimento di Scienze Fisiche Università di Napoli "Federico II" (Italy)*

[2] *COHERENTIA - INFM and Di.M.S.A.T Università di Cassino (Italy)*

[3] *Present address: IFF, Forschungszentrum Juelich, 52425 Juelich (Germany)*

**Abstract**

Scanning tunnelling spectroscopy measurements were performed on $La_{0.7}Sr_{0.3}MnO_3$ thin films both at room temperature and liquid nitrogen temperature. While no inhomogeneities were recorded at liquid nitrogen temperature on any sample, a clear evidence of spectroscopic inhomogeneities was evident in tunnelling conductance maps collected at room temperature. The investigated films exhibit a transition from a ferromagnetic-metallic to a paramagnetic-insulating state around room temperature, so that the observed spectroscopic features can be interpreted within a phase separation scenario. A quantitative analysis of the observed spectroscopic features is reported pointing out the occurrence of phase modulation and its possible correlation with the properties of the system.

Pacs 75.47.Lx, 71.30.+h, 87.64.Dz

## 1. Introduction

The rich phase diagram and the colossal magnetoresistance [1] of manganites such as $La_{1-x}A_xMnO_3$ (A being a divalent metallic atoms) strongly depend on the interplay between electron, spin, orbital and lattice degrees of freedom. It is widely accepted that, in many of these compounds for x close to x=0.3, the observed transition from a ferromagnetic metallic (FM) to a paramagnetic insulating (PI) state is due to the double-exchange mechanism proposed by Zener [2] together with a strong Jahn-Teller coupling responsible of the polaronic character of carriers at high temperatures [3,4]. Moreover many studies pointed out the possible formation of an inhomogeneous state (IS) with phase separation (PS) between FM and PI nanoscopic domains near the Curie temperature ($T_c$) [5,6]. While the existence of phase separation is supported by several theoretical approaches and experimental evidences [7-11], its characteristic length scale and dependence on external perturbations are still widely debated. Some recent experiments have found indications that even multi-scale phase modulation phenomena can occur in manganites [12].

Beside double-exchange and Jahn-Teller, several studies have stressed the role of disorder in inducing the phase coexistence [13], others proposed a metal-insulator transition induced by long-range elastic interactions that are known to be crucial in films [14,15] and bulk samples [16-18].

Within this framework, the Scanning Tunnelling Microscope (STM) represents a powerful tool to investigate the inhomogeneous state. Scanning Tunnelling Spectroscopy (STS) performed by STM differs from most other techniques, since it allows to image spectroscopic inhomogeneities, up to a nanometer resolution, in the real space. The direct observation of an inhomogeneous state in manganites near $T_c$ through STS is therefore possible because of the different spectroscopic features of the paramagnetic and ferromagnetic phases: insulating- and metallic-like respectively.

Differential tunnelling conductance can be recorded as a function of the position in the scanned area (dI/dV maps), so mapping the Local Density of States (LDOS) of the sample surface.

The presence of inhomogeneous state has been clearly observed on $La_{1-x}Ca_xMnO_3$ (LCMO). STS unambiguously revealed the presence of inhomogeneous structures and patterns in LDOS close to x=0.3 pointing out a submicrometer scale for inhomogeneities [8]. For what concerns $La_{1-x}Sr_xMnO_3$ (LSMO) the question is more controversial and puzzling. Although it has been suggested by some authors that the double exchange mechanism could fully describe the behaviour of LSMO bulk [19], results obtained with different techniques, such as resistivity data [20], optical conductivity [21], susceptibility measurements at low doping [22], photoemission and x-ray spectroscopy [23], provided indirect indications about the possible presence of an inhomogenus state even in bulk samples. STM measurements have not yet clarified the picture. To our knowledge, the only observation of electronic inhomogeneities on LSMO by STS was reported by Becker et al. [24], but they mainly reported on a different relative abundance of conducting and insulating spots as a function of temperature, and there was not a clear imaging of spatial features as for LCMO [8]. In contrast, Akiyama et al. [25] observed magnetic domains on LSMO thin films with a LSMO coated STM tip, but they measured homogeneous electronic LDOS with metallic tips.

In this paper, we report on directly imaged domains having, at room temperatures, different spectroscopic signatures, by performing STS measurements on $La_{0.7}Sr_{0.3}MnO_3$ thin films grown on different substrates. Considering the reproducibility of measurements, the electronic homogeneity systematically observed, in contrast, at liquid nitrogen temperature, and the absence of correlation between spectroscopy and topography, as discussed in the following, these results strongly suggest that a phase modulation can occur in LSMO when triggered by disorder and long-range strains.

In section 2 we briefly describe the films fabrication and main characterization, as well as the experimental procedure adopted for the STS measurements. Section 3 reports the observed tunnelling results: dI/dV maps on sample with different thicknesses at different temperatures, tunnelling spectra, STM topographies and basic comments on such results. Finally, in section 4 we perform a quantitative analysis and discuss our data and their interpretation.

## 2. Sample preparation and experimental details

$La_{0.7}Sr_{0.3}MnO_3$ thin films were fabricated by rf magnetron sputtering from stoichiometric target. Films stoichiometry was measured by energy dispersion spectroscopy (EDS) and Rutherford backscattering (RBS). Transport and magnetic characterizations are provided in more detail elsewhere [20,26].

Fig. 1 shows resistivity vs. temperature ($\rho(T)$) curves. The plotted curves refer to the same films, grown on $SrTiO_3$ (STO) substrates, for which STS measurements are reported. $T_p$ represents the temperature of maximum resistivity; $T_c$ values can be estimated from the maximum slope of $\rho(T)$ [5]. Table 1 summarizes the main properties of the films. We stress that, unlike bulk samples, LSMO films are characterized by a metal-insulator transition at x=0.3.

STM experiments were performed in inert helium atmosphere. The films were mounted on the STM scanner head and sealed in helium soon after the fabrication, limiting air exposure to preserve the surface quality. We used PtIr metallic tips, fabricated by an electrochemical etching procedure, which guarantees sharpness and reproducibility. The tips were tested by routinely achieving atomic resolution on graphite and $NbSe_2$ and flat conductance spectra on Au. A more detailed description of the experimental apparatus is reported in [27,28]. In the experiments on LSMO films, the junction quality was checked, according to what reported in [29], through the reproducibility of tunnel spectra vs. tunnel resistance and of topographic details without artefacts. These checks were performed before and after any measurement run.

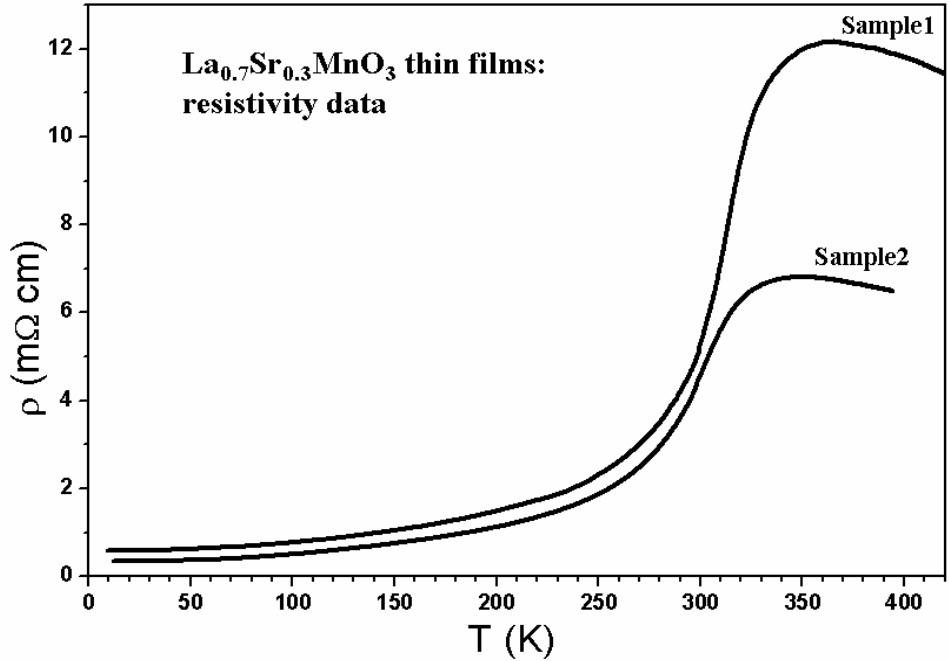

Fig. 1: ρ(T) curves for the investigated $La_{0.7}Sr_{0.3}MnO_3$ films. The transition temperature has been evaluated from these curves by taking the temperature of the maximum derivative [5].

TABLE 1: main properties of reported $La_{0.7}Sr_{0.3}MnO_3$ films.

| Name | Substrate | Thickness (nm) | $T_p$ (K) | $T_c$ (K) |
|---|---|---|---|---|
| Sample1 | STO (110) | 40 | 350 | 310 |
| Sample2 | STO (100) | 10 | 350 | 300 |

Topographic images were acquired in the constant current mode, while tunnelling spectra (tunnelling current and differential conductance vs. bias voltage curves) were recorded disconnecting the feedback loop and using a standard lock-in technique. Tunnelling conductance maps were imaged through Current Image Tunnelling Spectroscopy (CITS) measurements. In this technique, the tip is moved from a point to another, on a scan line, with the current kept constant by the feedback loop, as in the topographic mode. However, in each point, the feedback is disconnected in order to acquire the tunnelling conductance at a fixed bias voltage. Thus, a dI/dV map represents an image of the surface LDOS (whose distance from the Fermi level is determined by the bias) at a fixed energy; such an image is a powerful way to detect possible spectroscopic and electronic inhomogeneities on the sample. Different dI/dV maps on the same scan area were recorded to check the reproducibility of the imaged spectroscopic structures. The absence of artefacts was also tested by repeating the measurements at different scan rates. On the contrary, the same scan rate and the same experimental parameters were considered when the results on different samples had to be compared.

Acquisition of typical spectroscopic maps required about 13 hours to be completed. They were acquired both at room temperature and at 77 K. In the latter case, the thermal stability of the STM junction was guaranteed by the presence of liquid nitrogen. At room temperature, we experienced strong problems in achieving the stability required for such a long STM operation. For this reason, to avoid distortions in the tunnelling conductance maps, only when the STM topographic scans showed reproducibility over a long time we performed the spectroscopic measurements. Many days were often necessary before the dI/dV map could be acquired.

## 3. Experimental results

Fig. 2 summarizes the main observed features on Sample1 and Sample2. Tunnelling conductance maps are reported. In the maps, the lighter colour represents higher tunnelling conductance, i.e. a more metallic character of the surface. All dI/dV maps were measured at a tip to sample bias voltage of 2 V, with a tunnel current of 100 pA. As already reported in other STS measurements [8], this relatively high tunnel resistance is needed because of the poor metallic properties of the surface of these compounds. The maps in Fig. 2 were acquired at a scan rate of about 2.7 nm/s.

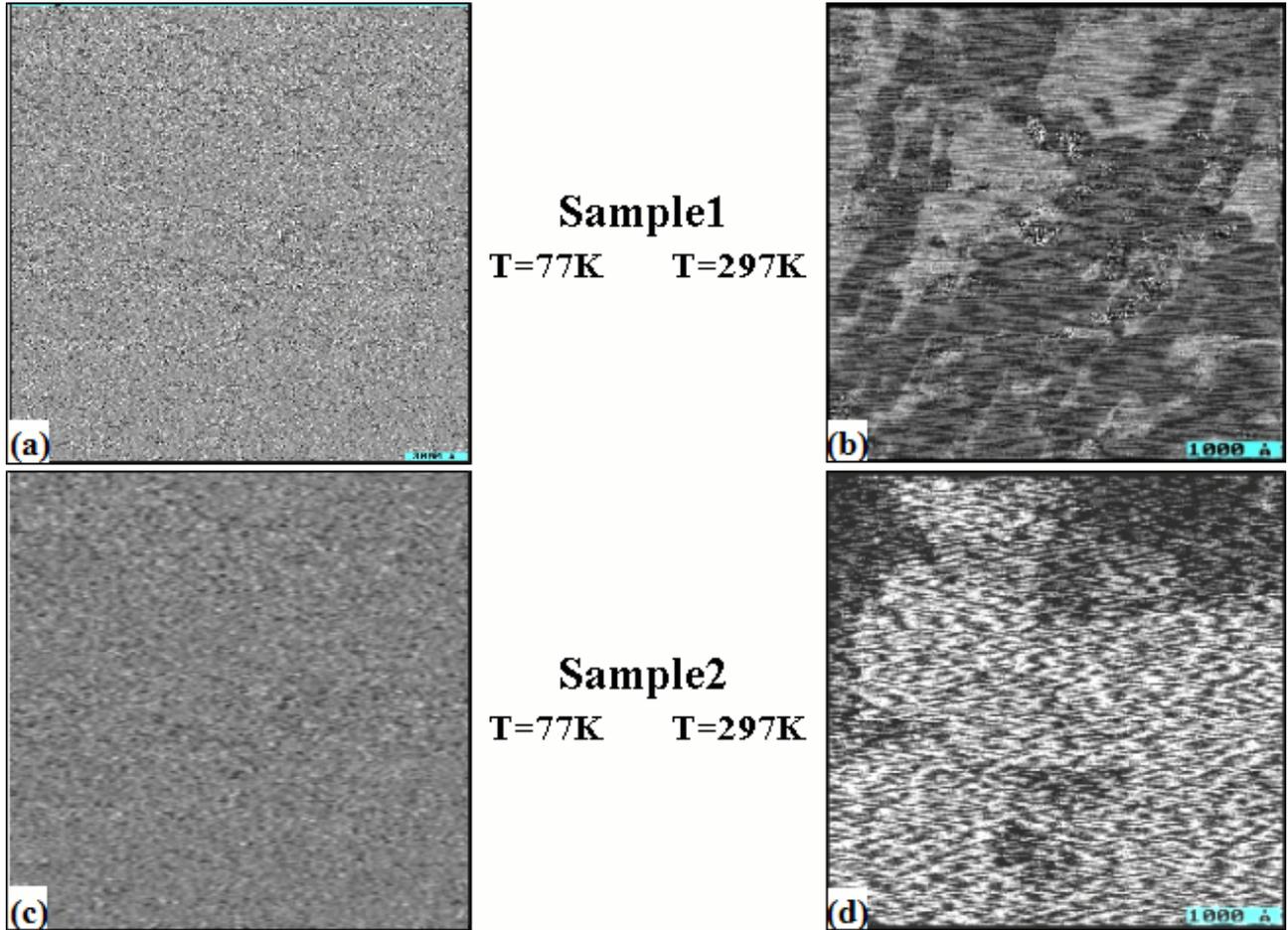

Fig. 2: dI/dV maps, 500x500nm$^2$: a) on Sample1 at T=77K and b) at T=297K; c) on Sample2 at T=77K and d) at T=297K. The maps were acquired at a scan rate of 2.7 nm/s, tunnel current of 100 pA, bias voltage (tip to sample) of 2V; in them, the lighter colours represent a more conducting character of the surface.

At 77 K, well below $T_c$ and therefore completely in the ferromagnetic state, dI/dV maps did not show evidence of inhomogeneities on any measured sample, exhibiting homogeneous LDOS on the whole imaged area (Figs. 2a and 2c). In contrast, maps at 297 K systematically showed submicrometric regions with sharply different spectroscopic features (Figs. 2b and 2d). From Table 1, we see that the measurement*s* temperature, 297 K, corresponds to a temperature slightly lower than $T_c$ for both samples. This circumstance agrees with most PS theoretical works, which predict that separation effects are expected in the temperature range around the transition, where insulating (paramagnetic) cluster could appear even below the transition temperature. Furthermore, we note that the tunnelling dI/dV values recorded at 77 K correspond to the more conducting state in the 297 K maps, as discussed in the following.

Films with two different thicknesses, 40 nm and 10 nm, were measured.

Tunnelling conductance maps on 40 nm thick films were imaged on samples grown on both STO(110) and STO(100): a dI/dV map acquired at room temperature on Sample1 is reported in Fig. 2b. Fig. 2d represents a dI/dV map at room temperature on Sample2, a 10nm thick film on STO(100) (good quality films with this thickness were only obtained on this substrate).

A clear appearance of "islands" in LDOS at room temperature was found on every sample at room temperature. The relative abundance of conducting and insulating zones in each map depends on the imaged scanned area; however, the typical lateral size of the "islands" ranges on a submicrometer scale, being of the order of 100-300 nm (depending on the single island).

A comparison between spectroscopic and topographic features has been performed, in order to check if spectroscopic and topographic features show some degree of correlation. Fig. 3a shows a topographic image on Sample2 on the same area of the dI/dV map (simultaneously acquired) reported in Fig. 2d.

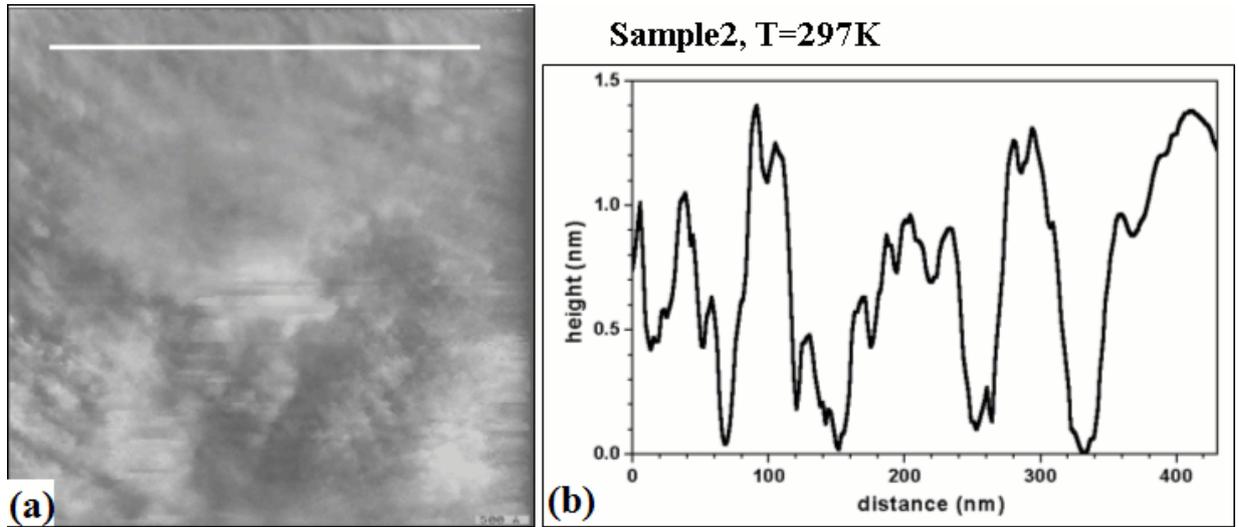

Fig. 3: a) STM topography on Sample2 at room temperature, on the same scanning area of the map in Fig. 2d; b) the height profile along the white line in the topography of Fig. 2d.

The data show a smooth surface: the height profile along the white line (crossing regions with different spectroscopic signatures), plotted in Fig. 3b, presents a peak to peak roughness lower than 15 Å over a range of more than 4000 Å. Such a surface represents therefore a good test to prove the absence of systematic correspondences between the two kinds of measurements. This claim can be quantitatively supported by evaluating the correlation coefficient of the two (spectroscopic and topographic) bidimensional values distributions:

$$R = \frac{\sigma_{ct}^2}{\sqrt{\sigma_c^2 \sigma_t^2}} = \frac{\sum_{ij} \left(\frac{C_{ij} - \overline{C}}{\overline{C}}\right)^2 \left(\frac{T_{ij} - \overline{T}}{\overline{T}}\right)^2}{\sqrt{\sum_{ij} \left(\frac{C_{ij} - \overline{C}}{\overline{C}}\right)^2 \sum_{ij} \left(\frac{C_{ij} - \overline{T}}{\overline{T}}\right)^2}}$$

In the previous formula, $C_{ij}$ and $T_{ij}$ are the conductance and topography values respectively, measured at each pixel, while $\overline{C}$ and $\overline{T}$ are their average values; $\sigma_{tc}^2$, $\sigma_c^2$, $\sigma_t^2$, are the covariance and the variances of the two distributions. The coefficient R ranges from 0 to 1, 1 representing the complete correlation between the two data sets. For the measurements reported in Figs. 2d and 3a, we calculated $R = 3 \times 10^{-2}$. It is worth to note that the topographic image used in this calculation was acquired simultaneously to the conductance map: during the CITS measurement the height value was recorded at each pixel before disconnecting the feedback. As a consequence, we can exclude any thermal drift effect to have produced displacements from a measure to another destroying a possible correlation.

In Fig. 4, two spectra acquired with the tip well within an insulating and a conducting island, respectively, are plotted. Both current and conductance curves reflect the different spectroscopic character of the two regions. The difference in the tunnelling spectra taken on surface portions with different conductance is not extremely marked: this well agrees with most reported tunnelling measurements on manganites.

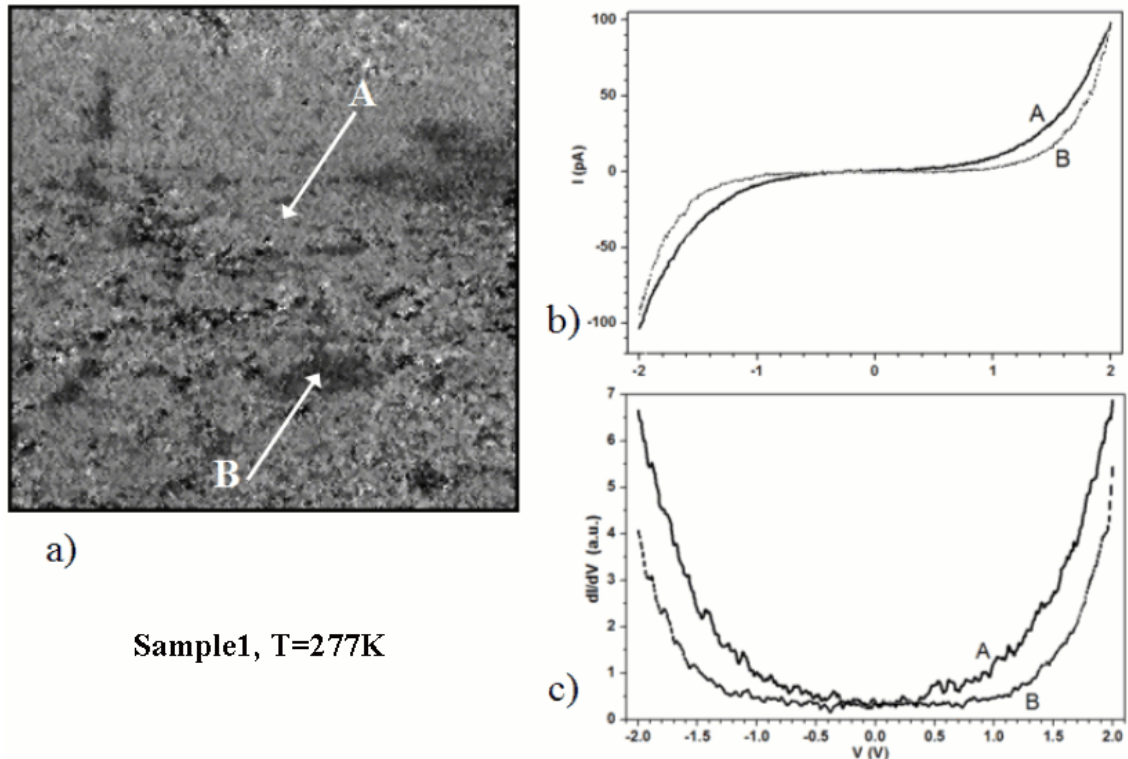

Fig. 4: a) dI/dV maps, 500x500nm$^2$, on Sample1 at T=297K; in the highlighted points A and B, tunnelling current and conductance vs. bias voltage curves were acquired; the curves are reported in b) (current) and c) (conductance).

The systematic observation of inhomogeneities just below $T_c$, the homogeneous appearance of all dI/dV maps at 77 K, the absence of correlation between dI/dV maps and topographic details, and the results in terms of tunnelling spectra, make us confident that the observed spectroscopic inhomogeneities are related to electronic and LDOS, intrinsic, spatial differences, and not connected to any accidental chemical inhomogeneity occurrence [5,12].

In addition, we should point out the presence, in the dI/dV maps, of a modulation in the tunnelling conductance on a smaller length scale (about 10-20 nm), more evident in Sample2. We do not yet have a clear explanation of this feature. This nanoscale pattern is present both in conducting and insulating submicrometric regions. At the moment we are not able to check if this feature is intrinsic or due to extrinsic effects. Some work in this direction is in progress.

We end this section mentioning that we observed both dI/dV patterns also after thermal cycles, and they were reproducible (including the nanometer scale modulation details) when changing the scan rate. This strongly suggests that the measured dI/dV features are not artefacts related to the tip-sample interaction. Since understanding and data analysis regarding this phenomenon are still at a preliminary stage, their discussion will not be addressed within this work

## 4. Analysis and discussion

A more quantitative analysis can be developed. Fig. 5 reports the dI/dV profile along scan lines in the maps recorded at room temperature on Sample1 and Sample2 (Figs. 4a and 4c), and the histogram distribution of the dI/dV values on the maps (Figs. 4b and 4d). All maps were recorded in the same conditions (described above), so a direct comparison between the values is possible. The histograms report, separately for conducting and insulating large islands, the relative counting frequency of the tunnel conductance values plotted in the dI/dV profiles. They represent a good synthesis of the information included in the maps and in the conductance line profiles. Indeed, the maps make evident the presence of inhomogeneities, but do not provide quantitative details on the tunnelling conductance values; on the other hand, the dI/dV line profiles allow an immediate comparison between spectroscopic measurements on different films and at different temperatures, but only on a very small portion of the experimental data. A histogram shows the quantitative content of a conductance profile, but representative of the whole area imaged in the map.

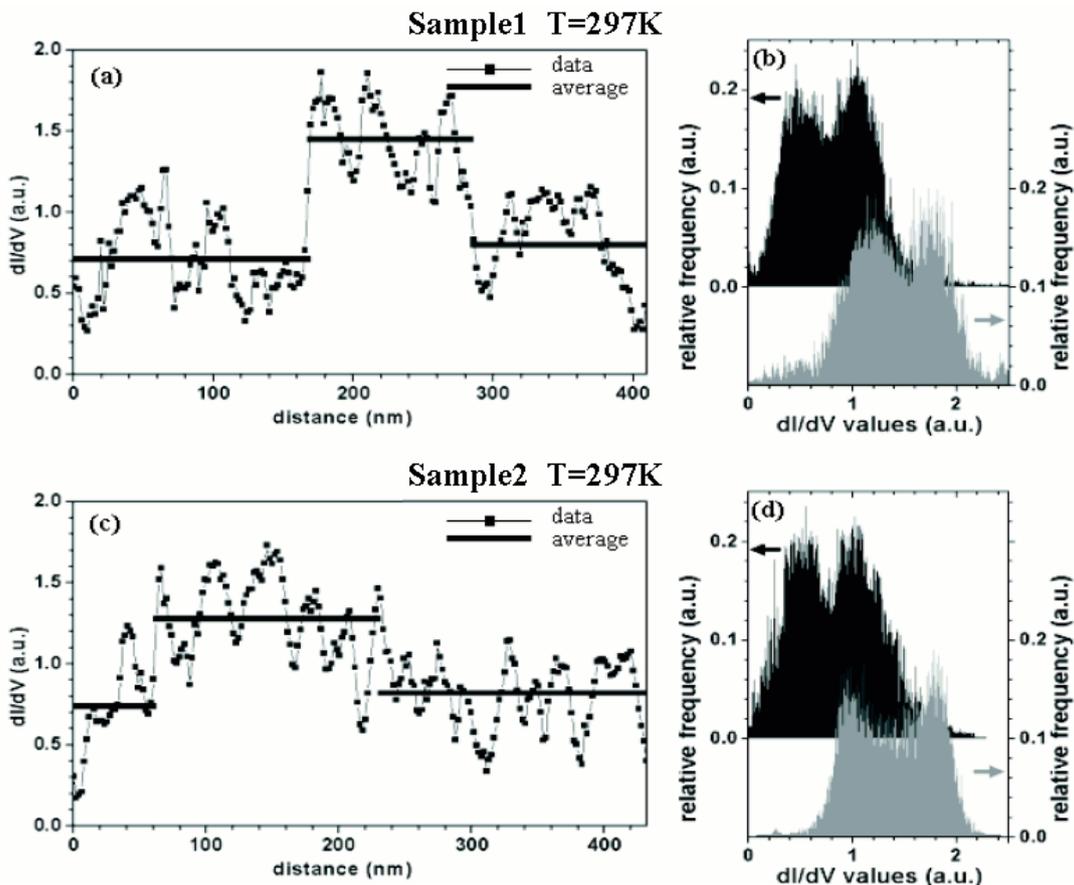

Fig. 5: STS room temperature maps analysis, concerning: a) dI/dV profile along a line for Sample1 (40 nm thick); b) dI/dV values distribution for Sample1; c) dI/dV profile along a line for Sample2 (10 nm thick); d) dI/dV values distribution for Sample2. Histograms have been plotted separately for the conducting and insulating large regions, and vertically shifted for clarity.

From the dI/dV profiles (Figs. 5a and 5c), the average values, taken on a line portion inside a "single-colour" region of the map, put in evidence the LDOS separation. The highest dI/dV value, i.e. the one corresponding to the more conducting regions, is very close to the one recorded on the whole map at 77 K (ranging between 1.3 and 1.5 in the adopted units). The lowest one (corresponding to the more insulating regions) is well reproducible from a sample to another and from a (less conducting) island to another. The two average values look clearly distinguishable, from both the profile and the histograms, resembling the sharpness of features shown by the maps. The double-peaked feature, as well as the periodic-like modulation in the line profiles are just a mere consequence of the above mentioned nanometric modulation in the dI/dV maps.

Several mechanisms have been proposed to explain the stabilization of submicrometric domains with different electronic and magnetic properties [13,14,15,16].

However, while for the LCMO, the coexistence of two phases around $T_c$, for x close to 0.3, is nowadays widely accepted the existence of inhomogeneities between ferromagnetic metallic and paramagnetic insulating regions in LSMO at x=0.3 is somehow unexpected. This is due to a more pronounced metallic character of LSMO than LCMO that makes not obvious the formation, below $T_c$, of insulating clusters resembling the phase above the transition.

Our results, that clearly support the existence of submicrometric domains in LSMO films, can be understood if we remember two experimental observations. First of all, Mannella et al. [23] have shown that LSMO single crystals exhibit different types of lattice distortions. These are smaller than in the LCMO case but, still, they could trigger a phase separation in LSMO. A second important observation is that, unlike single crystal, LSMO films at x=0.3 exhibit an insulating-like decrease of $\rho(T)$ above $T_P$ (see Fig.1). These circumstances suggest that LSMO thin films can be considered, under this point of view, more similar to the LCMO system than to LSMO single crystal. The difference between single crystal and films, in our opinion, can be interpreted in terms of the role played by disorder and strain in films. Actually, our observations are consistent with what reported by some of us in [20], where $\rho(T)$ curves for samples with different degree of disorder were analyzed with a model based on the PS scenario. Such analysis provided a good fit of the data for the samples having the highest residual resistivity and the less metallic behaviour at high temperature; it should be remarked that those samples exhibited $\rho_0$ values very similar to the one sof Sample1 and Sample2 reported in Fig. 1. The same model resulted in a poor fitting of data for a sample with lower resistivity, i.e. with a more metallic character. This occurrence suggests that disorder can play an important role in films favouring a metal insulating crossover and submicrometric phase coexistence, as proposed in [30]. It is worth to note that, alternatively to or in combination with disorder, the stress field induced by the substrate might play a major role in tuning the observed inhomogeneities [12].

Finally, as concerns the lower average dI/dV value observed on metallic regions of Sample2, it has the lowest $T_c$, and therefore it could be merely a consequence of the closest transition point. However, since the difference in the $T_c$'s is very small and the transitions quite broad, we cannot exclude an effect of the finite thickness [31] on this very thin film.

In conclusion, we provided clear, direct, observation of inhomogeneities in the surface LDOS on $La_{0.7}Sr_{0.3}MnO_3$ thin films by STS spectroscopy close to $T_c$. The main experimental finding is represented by the coexistence, in films grown on STO(110) and STO(100), of two phases with different electronic features just below the transition temperature from the insulating-paramagnetic to the metallic-ferromagnetic state. The same measurements well below $T_c$ show a homogenous aspect of the surface LDOS, which proves, as well as other discussed circumstances, that we observed an intrinsic properties of the measured sampled. Our data strongly suggest that the observed spectroscopic (electronic) pattern is related to the magnetic transition in the compound, being in agreement with the Phase Separation scenario predicted for manganites and still debated in the case of the LSMO system. The clearly observed islands on the sub-micrometer length scale, and in particular its apparent correlation with the disorder of the sample and with the substrate, seem to indicate the possibility to drive the Phase Separation by acting on strain and disorder. As concerns the observed conductance modulation on the smaller scale, although we cannot ignore its presence in the recorded dI/dV maps and related analysis, we believe that it requires a deeper investigation and measurements in order to clarify its presence and behaviour.

**References**

[1] S. Jin et al., Science 264, 413 (1994).
[2] C. Zener, Phys. Rev. 82, 403 (1951).
[3] A.J. Mills, P.B. Littlewood, and B.I. Shraiman, Phys. Rev. Lett. 74, 5144 (1995).
[4] M.B. Salamon and M. Jaime, Rev. Mod. Phys. 73, 583 (2001).
[5] E. Dagotto, Nanoscale Phase Separation and Colossal Magnetoresistance, Springer Berlin (2003).
[6] N. Mathur, P. Littlewood, Phys. Today Jan. 2003, 25.
[7] M. Uehara et al., Nature 399, 560 (1999).
[8] M. Fath et al., Science 285, 1540 (1999).
[9] Q. Lu, C.-C. Chen, and A. de Lozanne, Science 276, 2006 (1997).
[10] C.P. Adams et al., Phys. Rev. Lett. 85, 3954 (2000).
[11] S.A. McGill et al., Phys. Rev. Lett. 93, 47402 (2004).
[12] K.H. Kim et al., in Colossal Magnetoresistive Manganites, ed. T. Chatterji, Kluwer Academic Publishers (Dordrecht, Netherlands, 2002); also on cond-mat/0212113.
[13] E. Dagotto, T. Hotta, A. Moreo, Phys. Rep. 344, 1 (2001).
[14] Y. Konishi et al., J. Phys. Soc. Jpn. 68, 3790 (1999); Z. Fang, I.V. Solovyev, K. Terakura, Phys. Rev. Lett. 84, 3169 (2000).
[15] Y. Ogimoto, et al., Phys. Rev. B 71, 060403 (2005).
[16] M.J. Calderon, A.J. Millis, and K.H. Ahn, Phys. Rev. B 68, 100401 (2003).
[17] K.H. Ahn, T. Lookman, and A.R. Bishop, Nature 428, 401 (2004).
[18] D.D. Sarma, et al., Phys. Rev. Lett. 93, 97202 (2004).
[19] N. Furukawa, in Physics of Manganites, edited by Kaplan-Mahanti (Plenum, New York, 1999); Colossal Magnetoresistive Oxides, edited by Y. Tokura (Gordon and Breach Science Publishers, New York, 2000).
[20] S. Mercone et al., Phys. Rev. B 71, 064415 (2005).
[21] C. Hartinger, et al., cond-mat 0404023 (2004).
[22] J. Deisenhofer, et al., cond-mat/0501443.
[23] N. Mannella, et al., Phys. Rev. Lett. 92, 166401 (2004).
[24] T. Becker et al., Phys. Rev. Lett. 89, 237203 (2002).
[25] R. Akiyama et al., Appl. Phys. Lett. 79, 4378 (2001).
[26] U. Scotti di Uccio et al., submitted to Phys. Rev B., unpublished.
[27] M. Iavarone et al., Phys. Rev. B 65, 214506 (2002).
[28] M. Iavarone et al., Phys. Rev. B 71, 214502 (2005).
[29] Ch. Renner and O. Fisher, Phys. Rev. B 51, 9208 (1995).
[30] J. Burgy, A. Moreo, and E. Dagotto, Phys. Rev. Lett. 92, 97202 (2004).
[31] J. Z. Sun et al., Appl. Phys. Lett. 74, 3017 (1999).